# What Pakistani Computer Science and Software Engineering Students Think about Software Testing?

Luiz Fernando Capretz
*Western University*
*Department of Electrical & Computer Engineering*
London, Ontario, Canada – N6A5B9
lcapretz@uwo.ca

Abdul Rehman Gilal
*Universiti Teknologi PETRONAS*
*Department of Computer and Information Sciences*
Seri Iskander, Malaysia - 32610
rehman.gilal@utp.edu.my

*Abstract*—Software testing is one of the crucial supporting processes of the software life cycle. Unfortunately for the software industry, the role is stigmatized, partly due to misperception and partly due to treatment of the role. The present study aims to analyse the situation to explore what restricts computer science and software engineering students from taking up a testing career in the software industry. To conduct this study, we surveyed 88 Pakistani students taking computer science or software engineering degrees. The results showed that the present study supports previous work into the unpopularity of testing compared to other software life cycle roles. Furthermore, the findings of our study showed that the role of tester has become a social role, with as many social connotations as technical implications.

## I. INTRODUCTION

The role of tester does not appear as one of the preferred roles among the population of software developers, according to previous study results [1] [2]. Some studies point out the need for reversing people´s perceptions regarding this role [3] [4] by using career progression and other related mechanisms to reinforce the crucial contributions that a tester brings to the project.

However, when human aspects are not taken into account in software projects, an important piece of the puzzle for project staffing is overlooked. It has been pointed out that human and social aspects play a significant role in software testing practices [5] [6]. Attention to human factors in software testing in an academic setting has been encouraged by Capretz [7]. In a real-world environment, Santos et al. [8] found that software engineers with a positive attitude towards software testing can significantly influence those who have a negative attitude.

## II. METHODOLOGY

In this research, we studied the chances of software engineering and computer science students taking up software testing careers and their reasons. To that end, we conducted a survey of 88 Pakistani senior students taking computer science and software engineering degree at IBA University in Sukkur, Pakistan. The survey asked participants to share the advantages and disadvantages of pursuing a career as a software tester.

L.F. Capretz is also a visiting professor at Yale University-NUS College.

The survey asked three questions. The first two questions were open ended questions: 1) What are three PROs (in order of importance) of pursuing a career in software testing; and 2) What are three CONs (in order of importance) of pursuing a career in software testing. The third question asked participants to indicate their intentions of pursuing a career in software testing. They were given the option to answer with: "certainly not," "no," "maybe," "yes," and "certainly yes."

## III. RESULTS

The authors note that similar responses were merged, and duplicates were eliminated to ensure a better understanding and further analysis. The responses are summarized below in Table I, the PROs in Table II, and the CONs in Table III.

TABLE I. CHANCES TO TAKE UP TESTING CAREERS

| Responses (88) | Numbers | Percentage |
|---|---|---|
| Certainly Not | 13 | 15% |
| No | 14 | 16% |
| May be | 42 | 47% |
| Yes | 12 | 14% |
| Certainly Yes | 7 | 8% |
| **TOTAL** | **88** | **100%** |

We found 9 main PROs and 8 main CONs in total; these statements are listed respectively in Table II and Table III below. The most important reasons considered as PROs for taking up a testing career among the surveyed individuals are presented in Table II, along with their frequencies.

The most important reasons considered as PROs for taking up a testing career among the surveyed individuals are presented, along with their frequencies. The most frequent, with a 52% of respondents, shows the perception that the role of the tester provides many learning opportunities. This is followed by the belief of 43% of respondents pointing that testing tasks particularities make software tester a thinking job that requires critical analysis. The remaining two reasons with a 35% is that it is an important job and it is fun to break things.

In contrast, when asked about the CONs for taking up a testing career, respondents gave most importance to the following reasons: (a) it is a tedious and time consuming job for 41% of respondents; (b) it requires expertise and it is a difficult/complex task for 60%; and (c) 30% of the respondents pointed out career progression as an important impediment. The perception that other team members may become upset with the failures found by tester, thus becoming an anti-social role was picked by 18%. Lastly, 20% of the subjects noted that in the labor market the tester is a role for which salaries are lower than the average benefits and wages for other roles; that view is reinforced by the perception that software testers are treated as second-class citizens within a software project as indicated by 25% of the respondents.

TABLE II. PERCENTAGES OF SALIENT PROs

| PROs | Percentages |
|---|---|
| Learning opportunities | 52% |
| Important job | 35% |
| Easy job | 43% |
| Thinking job | 20% |
| More jobs | 18% |
| Monetary benefits | 17% |
| Suitable for "freshers" | 10% |
| Fun to break things | 35% |
| Increase product quality | 6% |

TABLE III. PERCENTAGES OF SALIENT CONs

| CONs | Percentages |
|---|---|
| Second-class citizen | 25% |
| Career development | 30% |
| Complexity/Expertise needed | 60% |
| Tedious/Time consuming | 41% |
| Prefer development | 26% |
| Less monetary benefits | 20% |
| Find others's errors | 18% |
| Learn nothing new | 10% |

IV. RESULTS AND DISCUSSIONS

The PROs show that subjects identify the role as a way for better approaching a new project when they are newcomers. This indicates that they see the opportunity to improve their soft skills and the need for creativity as a positive challenge for their careers. In addition, it was found that some respondents perceive the following two aspects as constructive: the presence of the role of tester through all project stages, and the fact that testing activities provide access to the full scope of the project— both modularization and integration strategy—in a short period of time. The authors believe then that respondents may perceive the role of tester as a professional growth opportunity. Nevertheless, further empirical studies to investigate this aspect of the study need to be conducted.

During the analysis of the CONs, it was noted that one of the most frequently cited Con reason was the possibility of team members becoming upset with the tester due to the review of the team members' builds. This could be due to the fact that testers may be accustomed to auditing and criticizing the work of others. Also, the preference for roles other than tester, due to their general acceptance, constitutes a conclusive statement regarding the unpopularity of the role of tester among respondents.

An overwhelming number of respondents picked the option 'No' or "Certainly Not" as responses. These results concur with prior studies [9] [10], which point to the tester role as one of the least popular roles among others, such as project manager, analyst, designer, programmer, and maintenance.

Meanwhile, the reasons supporting the 'Maybe' choice relate to the availability of better job offers. They also may reflect personal preferences and attraction to the role. Furthermore, it is the authors' belief that the tester is a role with more social connotations than technical inclinations, as reflected by the findings of the present investigation.

Nevertheless, software testing appears to be a neglected area in the software industry. The main reasons, among the subjects in the study, for taking or not taking up a testing career are strongly related to individual preferences and the availability of a job offer involving a more attractive or better-paid role. Also, respondents agree that the tester role offers an opportunity to know quickly what the project is all about because the individual has a role in all development stages and several automated tools that support the professional's performance of the role.

In summary, the present study confirms prior findings of the unpopularity of the role of tester, positioning the software tester role among those less favored by software students.